\documentclass[3p,times,twocolumn]{elsarticle}

\usepackage{ecrc}

\volume{00}

\firstpage{1}

\journalname{Nuclear Physics B Proceedings Supplement}

\runauth{A. Pich}

\jid{nuphbp}

\jnltitlelogo{Nuclear Physics B Proceedings Supplement}

\usepackage{amssymb}

\usepackage[figuresright]{rotating}

\biboptions{sort&compress}
\pdfmapfile{+txfonts.map}
\def\be{\begin{equation}}
\def\ee{\end{equation}}
\def\beqn{\begin{eqnarray}}
\def\eeqn{\end{eqnarray}}
\def\no{\nonumber}
\def\ba{\begin{array}{c}}
\def\bat{\begin{array}{cc}}
\def\ea{\end{array}}
\def\bi{\begin{itemize}}
\def\ei{\end{itemize}}

\def\cJ{{\cal J}}
\def\cO{{\cal O}}

\newcommand{\eqn}[1]{(\ref{#1})}
\newcommand{\bel}[1]{\be\label{#1}}

\begin{document}

\begin{frontmatter}


\title{Leptons and QCD}  
\author{Antonio Pich} 
\address{IFIC, Universitat de Val\`encia -- CSIC, Apt. Correus 22085, E-46071 Val\`encia, Spain} 

\begin{abstract}
Three important QCD-related aspects of the $\tau$ and $\mu$ dynamics are reviewed:
the determination of the strong coupling from the hadronic tau decay width,
leading to the updated value $\alpha_s(m_\tau^2) = 0.331 \pm 0.013$; the
measurement of $|V_{us}|$ through the Cabibbo-suppressed decays
of the $\tau$, and the Standard Model prediction for the muon anomalous magnetic moment.
\end{abstract}


\end{frontmatter}


\section{The $\tau$ hadronic width}
\label{intro}

The $\tau$ is the only known lepton massive enough to decay into hadrons.
Its semileptonic decays are then ideally suited to investigate the hadronic weak currents and perform low-energy tests of the strong interaction~\cite{Pich:2013lsa}.

The inclusive character of the $\tau$ hadronic width renders
possible an accurate calculation of the ratio
\cite{BR:88,NP:88,BNP:92,LDP:92a,QCD:94}
\be
 R_\tau \equiv { \Gamma [\tau^- \to \nu_\tau
 \,\mathrm{hadrons}] \over \Gamma [\tau^- \to \nu_\tau e^-
 {\bar \nu}_e] } \, =\,
 R_{\tau,V} + R_{\tau,A} + R_{\tau,S}\, .\;
\ee
The theoretical analysis involves the two-point correlation functions for
the vector $\, V^{\mu}_{ij} = \bar{\psi}_j \gamma^{\mu} \psi_i \, $
and axial-vector
$\, A^{\mu}_{ij} = \bar{\psi}_j \gamma^{\mu} \gamma_5 \psi_i \,$
colour-singlet quark currents ($i,j=u,d,s$; $\cJ = V,A$):
\be\label{eq:pi_v}
\Pi^{\mu \nu}_{ij,\cJ}(q) \equiv
 i \int d^4x \; e^{iqx}\,
\langle 0|T(\cJ^{\mu}_{ij}(x) \cJ^{\nu}_{ij}(0)^\dagger)|0\rangle \, ,
\ee
which have the Lorentz decompositions
\beqn\label{eq:lorentz}
\Pi^{\mu \nu}_{ij,\cJ}(q) & \!\!\! = & \!\!\!
  \left( -g^{\mu\nu} q^2 + q^{\mu} q^{\nu}\right) \: \Pi_{ij,\cJ}^{(1)}(q^2)
\no\\  &\!\!\! & \!\!\!
+\;   q^{\mu} q^{\nu} \, \Pi_{ij,\cJ}^{(0)}(q^2) \, ,
\eeqn
where the superscript $(J=0,1)$ denotes the angular momentum in the hadronic rest frame.

The imaginary parts of 
$\, \Pi^{(J)}_{ij,\cJ}(q^2) \, $
are proportional to the spectral functions for hadrons with the corresponding
quantum numbers.  The hadronic decay rate of the $\tau$
can be written as an integral of these spectral functions
over the invariant mass $s$ of the final-state hadrons:
\beqn\label{eq:spectral}
R_\tau  &\!\!\! = &\!\!\!
12 \pi \int^{m_\tau^2}_0 {ds \over m_\tau^2 } \,
 \left(1-{s \over m_\tau^2}\right)^2
\no\\ &\!\!\!\times &\!\!\!
 \biggl[ \left(1 + 2 {s \over m_\tau^2}\right)\,
 \mbox{\rm Im} \Pi^{(1)}(s)
 + \mbox{\rm Im} \Pi^{(0)}(s) \biggr] \, .
\eeqn
The appropriate combinations of correlators are
\beqn\label{eq:pi}
\Pi^{(J)}(s)  &\!\!\! \equiv  &\!\!\!
\sum_{q=d,s}\; |V_{uq}|^2 \, \left( \Pi^{(J)}_{uq,V}(s) + \Pi^{(J)}_{uq,A}(s) \right)\, .
\eeqn
The two terms with $q=d$ correspond to
$R_{\tau,V}$ and $R_{\tau,A}$ respectively, while
$R_{\tau,S}$ contains the remaining Cabibbo-suppressed contributions.

Since the spectral functions are sensitive to the non-perturbative effects that bind quarks into hadrons, the integrand in Eq.~(\ref{eq:spectral}) cannot be reliably predicted from QCD. Nevertheless the integral itself can be calculated 
by exploiting the fact that $\, \Pi^{(J)}_{ij,\cJ}(q^2) \, $ are analytic
functions of $s$, except along the positive real $s$-axis where their
imaginary parts have discontinuities.
Weighted integrals of the spectral functions can then be written as contour integrals in the complex $s$-plane running
counter-clockwise around the circle $|s|=m_\tau^2$ \cite{BNP:92,LDP:92b}:
\bel{eq:weighted_integrals}
\int_0^{s_0}\!\!\! ds\, w(s)\, \mathrm{Im}\Pi^{(J)}_{ij,\cJ}(s) =
\frac{i}{2} \oint_{|s|=s_0}\!\!\! ds\, w(s) \,\Pi^{(J)}_{ij,\cJ}(s) ,\;
\ee
with $w(s)$ an arbitrary weight function without singularities in the
region $|s|\leq s_0$.

The rhs of Eq.~\eqn{eq:weighted_integrals} requires the correlators only for
complex $s$ of order $s_0$. Provided $s_0$ is significantly larger than the scale associated with non-perturbative effects, one can use
the Operator Product Expansion (OPE),
$\Pi^{(J)}(s) = \sum_{D=2n} C_D^{(J)}/ (-s)^{D/2}$, to express
the contour integral as an expansion in powers of $1/m_\tau^2$ \cite{BNP:92}.
The $D=0$ term corresponds to the perturbative contribution, neglecting quark masses; non-perturbative physics appears at $D\ge 4$.
%
Several fortunate facts make $R_\tau$ particularly suitable for a precise theoretical analysis~\cite{BNP:92}:
\bi
\item[i)] The tau mass is large enough to safely use the OPE at $s_0=m_\tau^2$.

\item[ii)] The OPE is only valid in the complex plane, away from the
real axis where the physical hadrons sit. The contributions to the contour integral from the region near the real axis are heavily suppressed in $R_\tau$ by the presence in (\ref{eq:spectral}) of a double zero at $s=m_\tau^2$.

\item[iii)] For massless quarks,
$s \,\Pi^{(0)}(s) = 0$. Therefore, only the correlator
$\Pi^{(0+1)}(s)$ contributes to Eq.~(\ref{eq:spectral}), with a weight function $w(x) = (1 - x)^2 (1 + 2 x) =
1 - 3 x^2 + 2x^3$ [$x\equiv s/m_\tau^2$].
Cauchy's theorem guarantees that, up to tiny logarithmic running corrections, the only non-perturbative contributions to $R_\tau$ originate from operators of dimensions $D=6$ and 8. The usually leading $D=4$ operators can only contribute to $R_\tau$ with an additional suppression factor of $\cO(\alpha_s^2)$, which makes their effect negligible.

\item[iv)] While non-perturbative contributions to $R_{\tau,V}$ and $R_{\tau,A}$ are both suppressed by a factor $1/m_\tau^6$, the $D=6$ contributions to the vector and axial-vector correlators are expected to have opposite signs leading to a partial cancelation in $R_{\tau,V+A}$.
\ei

The theoretical prediction for the Cabibbo-allowed combination $R_{\tau,V+A}$
can be written as \cite{BNP:92}
\begin{equation}\label{eq:Rv+a}
 R_{\tau,V+A} \, =\, N_C\, |V_{ud}|^2\, S_{\mathrm{EW}} \left\{ 1 +
 \delta_{\mathrm{P}} + \delta_{\mathrm{NP}} \right\} ,
\end{equation}
where $N_C=3$ is the number of quark colours, $\delta_{\mathrm{NP}}$ the small non-perturbative contribution
and $S_{\mathrm{EW}}=1.0201\pm 0.0003$ contains the
electroweak corrections \cite{MS:88,BL:90,ER:02}.
The dominant effect is the perturbative QCD
contribution $\delta_{\mathrm{P}}\sim 20\%$.
%
The non-zero quark masses amount to a correction
smaller than $10^{-4}$  \cite{BNP:92,PP:99,BChK:05}.

The predicted value of $\delta_{\mathrm{P}}$ is very sensitive
to $\alpha_s(m_\tau^2)$, allowing for an accurate
determination of the fundamental QCD coupling \cite{NP:88,BNP:92}.
The calculation of the $\cO(\alpha_s^4)$ contribution \cite{BChK:08}
has triggered a renewed theoretical interest on the $\alpha_s(m_\tau^2)$ determination \cite{DDMHZ:08,Davier:2013sfa,Beneke:2008ad,Beneke:2012vb,Caprini:2011ya,Abbas:2012fi,Groote:2012jq,Maltman:2008nf,Boito:2012cr,Menke:2009vg,Narison:2009vy,Cvetic:2010ut,Pich:2011bb,Pich:2013sqa}, 
pushing the accuracy to the four-loop level.

\section{Perturbative contribution to $\boldmath R_\tau$}

The result is more conveniently expressed in terms of the
the logarithmic derivative of
$\Pi(s)=\frac{1}{2}\,\Pi^{(0+1)}(s)$,
which satisfies an homogeneous renormalization-group equation ($m_q=0$):
\be\label{eq:d}
 D(s)  \equiv
- s {d \over ds } \Pi(s)
=  {1\over 4 \pi^2} \sum_{n=0}  K_n
\left( {\alpha_s(-s)\over \pi}\right)^n \! .
\ee
%
For three flavours, the known coefficients take the values:
$K_0 = K_1 = 1$; $K_2 = 1.63982$; 
$K_3(\overline{MS}) = 6.37101$ and $K_4(\overline{MS}) =49.07570$  \cite{BChK:08}.

The perturbative component of $R_\tau$ is given by
\be\label{eq:r_k_exp}
\delta_{\mathrm{P}}\,  =\,
\sum_{n=1}  K_n \, A^{(n)}(a_\tau) \, =\, \sum_{n=1}\,  (K_n + g_n)\,
a_\tau^n \, ,
\ee
where $a_\tau\equiv\alpha_s(m_\tau^2)/\pi$ and
the contour integrals \cite{LDP:92a}
\beqn\label{eq:a_xi}
A^{(n)}(a_\tau) &\!\!\! = &\!\!\! {1\over 2 \pi i}\,
\oint_{|s| = m_\tau^2} {ds \over s} \,
  \left({\alpha_s(-s)\over\pi}\right)^n
\no\\ &\!\!\!\times  &\!\!\!
 \left( 1 - 2 {s \over m_\tau^2} + 2 {s^3 \over m_\tau^6}
         - {s^4 \over  m_\tau^8} \right)
\eeqn
%
can be numerically computed with high accuracy,
using the exact solution (up to unknown $\beta_{n>4}$ contributions) for $\alpha_s(-s)$ given by the renormalization-group $\beta$-function equation.
The resulting {\it contour-improved perturbation theory} (CIPT) series \cite{LDP:92a,PI:92}
has a very good perturbative convergence and is stable under changes of the renormalization scale.

One can instead expand the integrals in powers of $a_\tau$:
$A^{(n)}(a_\tau) = a_\tau^n + \cO(a_\tau^{n+1})$.
This procedure  \cite{BNP:92}, known as {\it fixed-order perturbation theory} (FOPT),
gives a rather bad approximation to $A^{(n)}(a_\tau)$, overestimating $\delta_{\mathrm{P}}$ by 12\% at $a_\tau = 0.11$ \cite{LDP:92a}.
The contour integration generates the $g_n$ coefficients which
are much larger than the original $K_n$ contributions \cite{LDP:92a}:
$g_1=0$, $g_2 =  3.56$, $g_3 = 19.99$, $g_4 = 78.00$, $g_5 = 307.78$.
FOPT suffers from a large renormalization-scale dependence \cite{LDP:92a};
its bad perturbative behaviour originates in the long running of $\alpha_s(-s)$ along the circle $|s|=m_\tau^2$ which makes compulsory to resum the large logarithms, $\log^n{(-s/m_\tau^2)}$, using the renormalization group \cite{LDP:92a}. This is precisely what CIPT does.

It has been argued that, once in the asymptotic regime (large $n$), the expected renormalonic behaviour of the $K_n$ coefficients could induce
cancelations with the running $g_n$ corrections, which would be missed by
CIPT. In that case, FOPT could approach faster the `true' result provided by the Borel summation of the full renormalon series.
Models of higher-order corrections with this behaviour have been advocated \cite{Beneke:2008ad}, but the results are however model dependent \cite{DM:10}.

\section{Determination of\ {\large $\mathbf{\alpha_s}$}}


The numerical size of $\delta_{\mathrm{NP}}$ can be determined from the measured invariant-mass distribution of the final hadrons in $\tau$ decay, through the study of weighted integrals which are more sensitive to OPE corrections~\cite{LDP:92b}.
%
%
The predicted suppression of $\delta_{\mathrm{NP}}$ has been confirmed by ALEPH \cite{ALEPH:05}, CLEO
\cite{CLEO:95} and OPAL \cite{OPAL:98}. The presently most complete and precise experimental analysis, performed with the ALEPH data, obtains
\cite{DDMHZ:08,Davier:2013sfa}
\begin{equation}\label{eq:del_np}
 \delta_{\mathrm{NP}} \, =\, -0.0064\pm 0.0013 \, .
\end{equation}
The QCD prediction for $R_{\tau,V+A}$ is then completely dominated
by $\delta_{\mathrm{P}}$;
non-perturbative effects
being smaller than the perturbative uncertainties.

Combining the $\tau$ lifetime and $e/\mu$ 
branching fractions into a {\it universality-improved} electronic branching ratio, the {\it Heavy Flavor Averaging Group} 
quotes~\cite{HFAG,Lusiani:2014aga}:
\beqn\label{eq:HFAGvalues}
R_{\tau,V+A} &\!\!\! =&\!\!\! 3.4696\pm 0.0080\, ,
\no\\
R_{\tau,S} &\!\!\! =&\!\!\! 0.1618 \pm 0.0026\, .
\eeqn
Using $|V_{ud}| = 0.97425\pm 0.00022$ \cite{PDG}
and Eq.~\eqn{eq:del_np}, the pure perturbative contribution to $R_\tau$ is determined to be:
\bel{eq:delta_P}
\delta_{\mathrm{P}} = 0.2009 \pm 0.0031 \, .
\ee

The main uncertainty in the $\tau$ determination of the strong coupling originates in the treatment of higher-order perturbative corrections~\cite{Pich:2013lsa}.
Using CIPT one gets $\alpha_s(m_\tau^2) = 0.341\pm 0.013$,
while FOPT would give
$\alpha_s(m_\tau^2) = 0.319\pm 0.014$.
Combining the two results, but keeping conservatively the smallest error, we get
\begin{equation}\label{eq:alpha-result}
\alpha_s^{(n_f=3)}(m_\tau^2)\; =\; 0.331 \pm 0.013\, .
\end{equation}

The direct analysis of the ALEPH invariant-mass distribution \cite{DDMHZ:08,Davier:2013sfa} determines
$\alpha_s(m_\tau^2)$ and the ($D\le 8$) OPE corrections through a global fit of $R_{\tau,V+A}$ and four weighted integrals [Eq.~\eqn{eq:weighted_integrals}] with $s_0=m_\tau^2$ and weights
$w(x) = (1+2x) (1-x)^3 x^l$ ($x=s/m_\tau^2$, $l=0,1,2,3$). Using CIPT, this gives $\delta_{\mathrm{NP}}$ in Eq.~\eqn{eq:del_np} and
$\alpha_s(m_\tau^2) = 0.341\pm 0.008$, fully consistent with our CIPT result.

The value of $\alpha_s$ in Eq.~\eqn{eq:alpha-result} is
significantly larger ($16\,\sigma$) than the result obtained from the $Z$ hadronic width,
$\alpha_s^{(n_f=5)}(M_Z^2) = 0.1197\pm 0.0028$ \cite{PDG} ($n_f$ denotes the relevant number of quark flavours at the given energy scale).
After evolution up to the scale $M_Z$ \cite{Rodrigo:1997zd,vanRitbergen:1997va,Czakon:2004bu,Schroder:2005hy,Chetyrkin:2005ia}, the strong coupling in \eqn{eq:alpha-result} decreases to $\alpha_s^{(n_f=5)}(M_Z^2) = 0.1200\pm 0.0015$,
in excellent agreement with the direct measurement at the $Z$ peak.
The comparison of these two determinations
provides a beautiful test of the predicted QCD running;
{\it i.e.}, a very significant experimental verification of asymptotic freedom:
\be
\left.\alpha_s(M_Z^2)\right|_\tau - \left.\alpha_s(M_Z^2)\right|_Z
\, = \, 0.0003 \pm 0.0032\, .
\ee

\section{Duality violations}

When the OPE is used to perform the contour integration \eqn{eq:weighted_integrals}, one is neglecting the difference
$\Delta_{ij,{\cal J}}^{(J)}(s) \equiv \Pi_{ij,{\cal J}}^{(J)}(s) -
\Pi_{ij,{\cal J}}^{(J),\mathrm{OPE}}(s)$. The missing correction
can be expressed as \cite{CGP:09,Martin}
\beqn\label{eq:duality-violation}
\frac{i}{2} \oint_{|s|=s_0}\!\!\! ds\, w(s) \,\Delta^{(J)}_{ij,\cJ}(s)
 = -\int^\infty_{s_0}\!\!\! ds\, w(s)\, \mathrm{Im}\,\Delta^{(J)}_{ij,\cJ}(s)\, .
\no
\eeqn
This effect is negligible in $R_\tau$, since it is smaller than the errors induced by $\delta_{\mathrm{NP}}$ which are in turn subdominant with respect to the leading perturbative uncertainties; however, it could be more relevant for other weighted integrals of the invariant-mass distribution.

Parametrizing $\mathrm{Im}\,\Delta^{(J)}_{ij,\cJ}(s)$ with the ansatz \cite{CGP:09,Shifman:2000jv}
\bel{eq:DVansatz}
\mathrm{Im}\,\Delta^{(J)}_{ij,\cJ}(s)\, =\, \pi\; \mathrm{e}^{-(\delta+\gamma s)}\,\sin{(\alpha + \beta s)}\, ,
\ee
the $\tau$ data can be used to fit the parameters $\alpha,\beta,\gamma,\delta$, which are different for each correlator $\Pi^{(J)}_{ij,\cJ}$.
In order to maximize duality violations, Refs.~\cite{Boito:2012cr} analyze the weight $w(x)\! =\! 1$ and fit the $s_0$ dependence of the corresponding $V$ and $A$ integrated distributions in the range $s_{\mathrm{min}}\!\equiv\! 1.55\:\mathrm{GeV}^2\!\le\! s_0\!\le\! m_\tau^2$. This is equivalent to a direct fit of the measured spectral functions in this energy range,\footnote{
The derivative of the integral of the spectral function is obviously the spectral function itself.}
plus the total integral at $s\! =\! s_{\mathrm{min}}$. Thus, one pays a very big price because
i) $\sqrt{s_{\mathrm{min}}}\! =\! 1.2\:\mathrm{GeV}$ is too low to be reliable; ii) one directly touches the real axis where the OPE is not valid, and iii) the separate $V$ and $A$ correlators have larger non-perturbative contributions than $V+A$. In addition, one has a too large number of free parameters to be fitted to a highly correlated data set.
In spite of all these caveats, this procedure results in quite reasonable
values of the strong coupling (CIPT): $\alpha_s(m_\tau^2) = 0.310\pm 0.014 \;\mathrm{(ALEPH)},\; 0.322\pm 0.026 \;\mathrm{(OPAL)}$~\cite{Boito:2012cr}. Although the quoted uncertainties seem underestimated, this suggests a much better behaviour of perturbative QCD at low values of $s$ than naively expected. This had been already noticed longtime ago in the pioneering analyses of the $s_0$ dependence performed in Refs.~\cite{ALEPH:05,Narison:1993sx,Girone:1995xb}.

The violations of quark-hadron duality
could play a more important role in observables which are not dominated by the perturbative contribution. A gold-plated example is
$\Pi_{LR}(s) = \Pi^{(0+1)}_{ud,V}(s) - \Pi^{(0+1)}_{ud,A}(s)$ which vanishes identically to all orders of perturbation theory. The $\tau$-data analysis of this correlator has allowed us to extract important information on
low-energy couplings of Chiral Perturbation Theory and other non-perturbative QCD parameters~\cite{GonzalezAlonso:2010xf,Boito:2012nt,Boyle:2014pja,Davier:1998dz}.

\section{$\mathbf{V_{us}}$ determination}

The separate measurement of the $|\Delta S|\! =\! 1$ and $|\Delta S|\! =\! 0$
tau decay widths provides a very clean determination of $V_{us}\,$
\cite{GJPPS:05,Gamiz:2013wn}. To a first approximation, their ratio directly measures $|V_{us}/V_{ud}|^2$. The experimental values in Eq.~\eqn{eq:HFAGvalues} imply
$|V_{us}|^{\mathrm{SU(3)}}\! =\! 0.210\pm 0.002$.
This rather remarkable determination is only slightly shifted by
the small SU(3)-breaking contributions induced by the strange quark mass.
These effects can be theoretically
estimated through a careful QCD analysis of the difference
\cite{GJPPS:05,Gamiz:2013wn,PP:99,ChDGHPP:01,ChKP:98,KKP:01,MW:06,KM:00,MA:98,BChK:05}
\begin{equation}
 \delta R_\tau  \,\equiv\,
 {R_{\tau,V+A}\over |V_{ud}|^2} - {R_{\tau,S}\over |V_{us}|^2}\, .
\end{equation}
%
The only non-zero contributions are proportional to the mass-squared difference $m_s^2-m_d^2$ or to vacuum expectation
values of SU(3)-breaking operators such as $\delta O_4
\equiv \langle 0|m_s\bar s s - m_d\bar d d|0\rangle \approx (-1.4\pm 0.4)
\cdot 10^{-3}\; \mathrm{GeV}^4$ \cite{PP:99,GJPPS:05}. The dimensions of these operators
are compensated by corresponding powers of $m_\tau^2$, which implies a strong
suppression of $\delta R_\tau$ \cite{PP:99}:
\beqn\label{eq:dRtau}
 \delta R_\tau &\!\!\approx &\!\!  24\, S_{\mathrm{EW}}\; \left\{ {m_s^2(m_\tau^2)\over m_\tau^2} \,
 \left( 1-\epsilon_d^2\right)\,\Delta_{00}(\alpha_s)
 \right.\no\\ &&\hskip 1.3cm\left.
 - 2\pi^2\, {\delta O_4\over m_\tau^4} \, Q_{00}(\alpha_s)\right\}\, ,
\eeqn
where $\epsilon_d\equiv m_d/m_s = 0.053\pm 0.002$ \cite{LE:96}.
The perturbative 
corrections $\Delta_{00}(\alpha_s)$ and
$Q_{00}(\alpha_s)$ are known to $O(\alpha_s^3)$ and $O(\alpha_s^2)$,
respectively \cite{PP:99,BChK:05}.

The $J=0$ contribution to $\Delta_{00}(\alpha_s)$ shows a rather
pathological behaviour, with clear signs of being a non-convergent perturbative
series. Fortunately, the corresponding longitudinal contribution to
$\delta R_\tau$ can be estimated phenomenologically with good
accuracy, $\delta R_\tau|^{L}\, =\, 0.1544\pm 0.0037$ \cite{GJPPS:05},
because it is dominated by far by the well-known $\tau\to\nu_\tau\pi$
and $\tau\to\nu_\tau K$ contributions.
To estimate the remaining transverse component, one needs an input value for the strange quark mass; we adopt the lattice world average \cite{Aoki:2013ldr}, but increasing conservatively its uncertainty to 6 MeV, {\it i.e.}
$m_s^{\overline{\mathrm{MS}}}(2~\mathrm{GeV}) = (94\pm 6)~\mathrm{MeV}$. The numerical size of $\Delta_{00}^{(0+1)}(\alpha_s)$ is estimated in a very conservative way, averaging the asymptotically summed CIPT and FOPT results and taking half of the difference as the uncertainty associated with the truncation of the series. One gets in this way $\delta R_{\tau,th} = 0.240\pm 0.032$ \cite{GJPPS:05,Gamiz:2013wn}, which implies
\beqn\label{eq:Vus_det}
 |V_{us}| &=& \left(\frac{R_{\tau,S}}{\frac{R_{\tau,V+A}}{|V_{ud}|^2}-\delta
 R_{\tau,\mathrm{th}}}\right)^{1/2}
 \no\\ &=&
  0.2177\pm 0.0018_{\mathrm{\, exp}}\pm 0.0010_{\mathrm{\, th}}\, .
\eeqn
This result is lower than the most recent
determination from $K_{\ell 3}$ decays, $|V_{us}|= 0.2229\pm 0.0009$ \cite{Cirigliano:2011ny,Pich:2014zta}.

The $\tau$ branching ratios measured by BaBar and Belle are smaller than previous
world averages, which translates into smaller results for $R_{\tau,S}$ and $|V_{us}|$. The measured $K^-\!\to\!\mu^-\bar\nu_\mu$ decay width implies a $\tau^-\!\to\!\nu_\tau K^-$ branching ratio $1.7\,\sigma$ higher than the present experimental value \cite{Pich:2013lsa}.
Combining the measured spectra in $\tau^-\to\nu_\tau (K\pi)^-$ decays with $K_{\ell 3}$ data \cite{Antonelli:2013usa}, one also predicts $\tau^-\to\nu_\tau \bar K^0\pi^-$ and $\tau^-\to\nu_\tau K^-\pi^0$ branching ratios $1.0\,\sigma$ and $1.6\,\sigma$ higher, respectively, than the world averages.
Replacing the direct $\tau$ decay measurements by these phenomenological estimates, one gets the corrected result
$R_{\tau,S} = 0.1665\pm 0.0034$ \cite{Pich:2013lsa}, which implies
\be
|V_{us}| = 0.2208\pm 0.0025\, ,
\ee
in much better agreement with the $K_{\ell 3}$
value.
Contrary to $K_{\ell 3}$, the final error of the $V_{us}$ determination from
$\tau$ decay is dominated by the experimental uncertainties and, therefore, sizeable improvements can be expected.

\section{Anomalous magnetic moments}
\label{sec:g-2}

The most stringent QED test \cite{Jegerlehner:2009ry,Miller:2012opa}
comes from the high-precision
measurements of the $e$ \cite{Hanneke:2008tm} 
and $\mu$ \cite{Bennett:2006fi}
anomalous magnetic moments, $a_l\equiv (g^\gamma_l-2)/2$:
\beqn\label{eq:a_e}
 a_e & =& (1\; 159\; 652\; 180.73\pm 0.28) \,\cdot\, 10^{-12}\, ,
 \no\\ \label{eq:a_mu}
 a_\mu &=& (11\; 659\; 208.9\pm 6.3) \,\cdot\, 10^{-10}\, .
\eeqn
The $O(\alpha^5)$ calculation has been completed in both cases
\cite{Aoyama:2012wj}, with an impressive agreement with the measured $a_e$ value.
The dominant QED uncertainty is the input value of $\alpha$, therefore
$a_e$ provides the most accurate determination of the fine structure constant (0.25 ppb),
\bel{eq:alpha}
\alpha^{-1} = 137.035\; 999\; 174 \,\pm\, 0.000\; 000\; 035\, ,
\ee
in agreement with the next most precise value (0.66 ppb)
$\alpha^{-1}_{\mathrm{Rb}} = 137.035\, 999\, 037 \pm 0.000\, 000\, 091$
\cite{Bouchendira:2010es}, deduced from the measured ratio $h/m_{\mathrm{Rb}}$ between the Planck constant and the mass of the ${}^{87}$Rb atom.
The improved experimental accuracy on $a_e$
is already sensitive to the hadronic contribution $\delta a_e^{\mathrm{QCD}}= (1.685\pm 0.033)\times 10^{-12}$, and approaching the level of the weak correction $\delta a_e^{\mathrm{ew}}= (0.0297\pm 0.0005)\times 10^{-12}$~\cite{Aoyama:2012wj}.

\begin{figure}[tbh]
\begin{center}
\includegraphics[width=\linewidth]{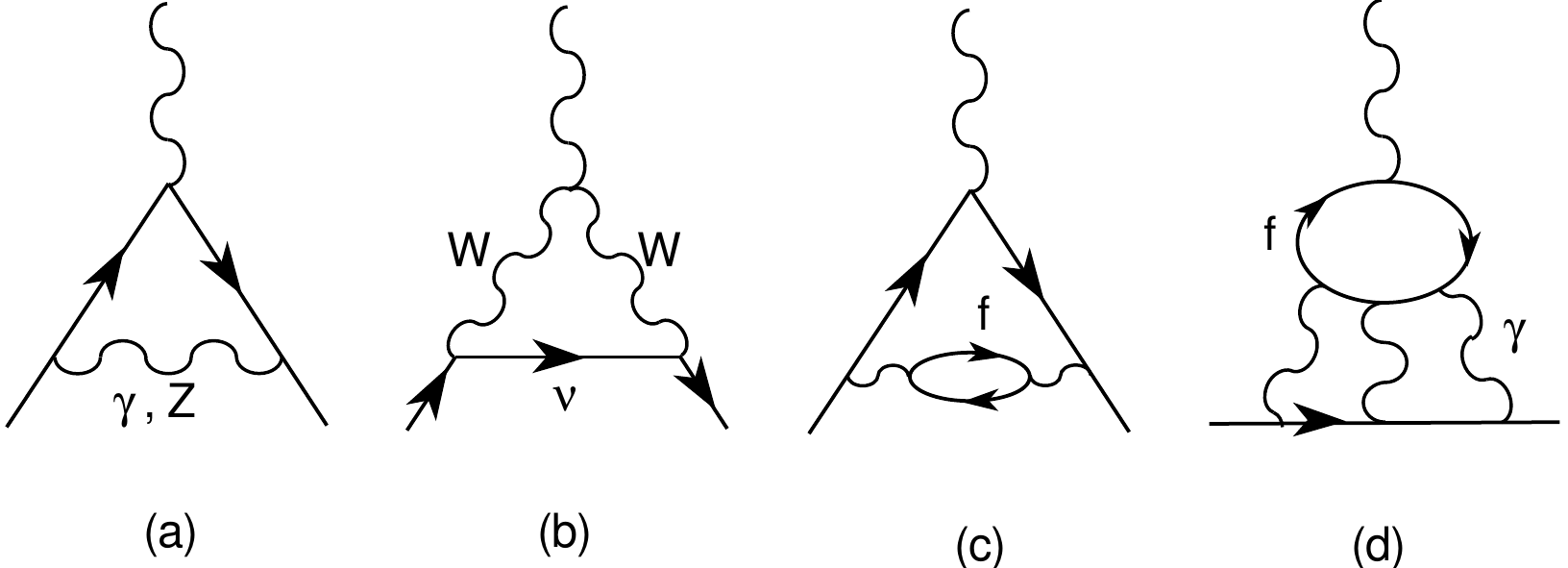}
\caption{Feynman diagrams contributing to the lepton anomalous magnetic moment.}
\label{fig:AnMagMom}
\end{center}
\end{figure}

The heavier muon mass makes $a_\mu$ much more sensitive to electroweak
corrections \cite{Czarnecki:1995wq,Czarnecki:2001pv,Czarnecki:2002nt,Gribouk:2005ee,Knecht:2002hr,Heinemeyer:2004yq,Kukhto:1992qv,Gnendiger:2013pva}
from virtual heavier states; compared to $a_e$, they scale as $m_\mu^2/m_e^2$.
The main theoretical uncertainty
comes from the
{\it hadronic vacuum polarization} corrections to the photon propagator
(Fig.~\ref{fig:AnMagMom}c),
which cannot be calculated at present with the required precision and must be extracted
\cite{Davier:2010nc,Jegerlehner:2009ry,Hagiwara:2011af,Jegerlehner:2011ti}
from the measurement of
$\sigma(e^+e^-\to \mathrm{hadrons})$ and from the
invariant-mass distribution of the final hadrons in $\tau$ decays.
$\delta a_\mu^{\mathrm{hvp,LO}}$ is dominated by the low-energy spectral region; the largest contribution being the $2\pi$ final state.
There is still a slight discrepancy between the $2\pi$ spectral functions extracted from $e^+e^-$ and $\tau$ data, which cannot be accounted for through isospin-breaking corrections
\cite{Cirigliano:2002pv,Davier:2009ag,Davier:2010nc,FloresBaez:2006gf,Davier:2003pw,Alemany:1997tn}.
Additional disagreements among $e^+e^-$ experiments remain in several final states \cite{Davier:2013-ARNPS}.

The Standard Model prediction for $a_\mu$ can be decomposed in five types of contributions:
\beqn 10^{10} \times a_\mu^{\mathrm{th}}\;  =\;
11\; 658\; 471.895 \pm 0.008 && \mathrm{QED}
\nonumber\\ \mbox{}
+\phantom{6}15.4\phantom{95}\pm 0.1\phantom{08} && \mathrm{EW}
\nonumber\\ \mbox{}
+ 697.4\phantom{95} \pm 5.3\phantom{08}
&& \mathrm{hvp}^{\mathrm{LO}}
\nonumber\\ \mbox{}
-\phantom{69}8.6\phantom{95}\pm 0.1\phantom{08}&& \mathrm{hvp}^{\mathrm{NLO}}
\nonumber\\ \mbox{}
+\phantom{6}10.5\phantom{95}\pm 2.6\phantom{08} &&\mathrm{lbl}
\nonumber\\
= \; 11\; 659\; 186.6\phantom{95} \pm  5.9\phantom{08} &&\hskip -1.cm .
\eeqn
The first line gives the QED contribution, 
including the recently computed $O(\alpha^5)$ corrections \cite{Aoyama:2012wj}. The quoted number adopts as input the value of $\alpha$ determined from the ${}^{87}\mathrm{Rb}$ atom; using instead the $a_e$ determination in \eqn{eq:alpha}, one gets the slightly more precise
result $\delta a_\mu^{\mathrm{QED}} = (11\, 658\, 471.885\pm 0.004)\times 10^{-10}$
\cite{Aoyama:2012wj}. The electroweak correction  \cite{Czarnecki:1995wq,Czarnecki:2001pv,Czarnecki:2002nt,Gribouk:2005ee,Knecht:2002hr,Heinemeyer:2004yq,Kukhto:1992qv,Gnendiger:2013pva} is shown in the second line.
The leading-order hadronic-vacuum-polarization contribution in the third line is a weighted average of the $\tau$ and $e^+e^-$ determinations:
$10^{10}\times \delta a_\mu^{\mathrm{hvp,LO}} = (703.0\pm 4.4)_\tau \, ,\,
(692.3\pm 4.2)_{e^+e^-}$
\cite{Davier:2010nc,Davier:2013sfa}. The sum of the estimated next-to-leading
\cite{Hagiwara:2011af,Krause:1996rf} and next-to-next-to-leading \cite{Kurz:2014wya} corrections is given in the fourth line.
Additional QCD uncertainties stem from the smaller {\it light-by-light scattering} contribution (Fig.~\ref{fig:AnMagMom}.d),
given in the fifth line \cite{Prades:2009tw}, which needs to be theoretically evaluated
\cite{Prades:2009tw,Bijnens:2007pz,Melnikov:2003xd,Knecht:2001qg,deRafael:1993za,Greynat:2012ww,Blokland:2001pb,Bijnens:2001cq,Hayakawa:1997rq,RamseyMusolf:2002cy,Goecke:2010if,Blum:2013qu,Colangelo:2014pva,Pauk:2014rta,Roig:2014uja,Benayoun:2014tra,Czyz:2013zga}.

The final Standard Model prediction differs from the experimental value by $2.6\,\sigma$. The $\tau$ estimate of the hadronic vacuum polarization results in a smaller deviation of $2.1\,\sigma$,
$a_\mu^{\mathrm{th}}|_\tau = (11\, 659\, 192.2\pm  5.1)\times 10^{-10}$,
while using $e^+e^-$ data alone increases the discrepancy to $3.4\,\sigma$,
$a_\mu^{\mathrm{th}}|_{e^+e^-} = (11\, 659\, 181.5\pm  4.9)\times 10^{-10}$.
Improved theoretical predictions and more precise $e^+e^-$ and $\tau$ data sets are needed to settle the true value of $a_\mu^{\mathrm{th}}$ and match the aimed $10^{-10}$ accuracy of the forthcoming muon experiments at Fermilab and J-PARC \cite{FNAL-JPARC}.

With a predicted value $a_\tau^{\mathrm{th}}= 117\, 721\, (5)\cdot 10^{-8}$ \cite{Eidelman:2007sb}, the $\tau$ anomalous magnetic moment has an enhanced sensitivity to new physics because of the large tau mass. However, it is essentially unknown experimentally:
$a_\tau^{\mathrm{exp}}= -0.018\pm 0.017$ \cite{Abdallah:2003xd}.
Using an effective Lagrangian, invariant under the Standard Model gauge group, and writing the lowest-dimension ($D=6$) operators contributing to $a_\tau$, it is possible to combine experimental information from $\tau$ production at LEP1, LEP2 and SLD with $W^-\to\tau^-\bar\nu_\tau$ data from LEP2 and $p\bar p$ colliders. This allows one to set a stronger model-independent bound on new-physics contributions to $a_\tau$ (95\% CL) \cite{GonzalezSprinberg:2000mk}:
\bel{eq:atau-new}
- 0.007\; <\; a_\tau^{\mathrm{New\, Phys}}\; <\; 0.005\, .
\ee
%


\section*{Acknowledgements} 
I would like to thank Achim Stahl and Ian M. Nugent for organizing this interesting conference, and Mart\'{\i}n Gonz\'alez-Alonso for his comments on the manuscript. This work has been supported by the Spanish Government~[grants FPA2011-23778 and CSD2007-00042] and the Generalitat Valenciana [PrometeoII/2013/007].


\end{document}